\documentclass[twocolumn]{revtex4}

\usepackage{bbm}
\usepackage{epsf}
\usepackage{amsmath}
\usepackage{amssymb}

\newcommand{\be}{\begin{equation}}
\newcommand{\ee}{\end{equation}}
\newcommand{\bea}{\begin{eqnarray}}
\newcommand{\eea}{\end{eqnarray}}
\newcommand{\R}{O}

\renewcommand{\v}[1]{\mathbf{#1}}

\begin{document}

\title{Evolutionary Minority Games: the benefits of imitation}
\author{Richard Metzler}
\affiliation{Institut f\"{u}r Theoretische Physik und
  Astrophysik, Universit\"{a}t W\"{u}rzburg, Am Hubland,
  D-97074 W\"{u}rzburg, Germany}
\author{Christian Horn}
\affiliation{ Experimentalphysik III, Universit\"{a}t
  Kassel, Heinrich-Plett-Str.40, D-34132 Kassel, Germany}

\begin{abstract}
In the original Evolutionary Minority Game, a segregation
into two populations with opposing preferences is observed
under many circumstances. We show that this 
segregation becomes more pronounced and more robust if
the dynamics are changed slightly, such that strategies with
above-average fitness become more frequent. Similar effects
occur also for a generalization of the EMG to more than two
choices, and for evolutionary dynamics of a different
stochastic strategy for the Minority Game.
\end{abstract}

\maketitle
\section{Introduction}
The Minority Game (MG) was introduced 
by Challet and Zhang in 1997 \cite{Challet:Emerg.}
as a model for the competition for limited resources.
Although it has since been studied in more than 100 publications
\cite{MGHomepage}, and countless variations have been 
introduced, the basic scenario is still easy to explain:
there is a population of $N$ players who, at each time step $t$,
have to make a decision $\sigma_i^t \in \{-1,1\}$. 
Those who are in the minority win, the others lose
(to avoid ambiguities, $N$ is chosen to be odd).
Direct communication and contracts among players are 
not allowed; however, the decision of the minority is
public information, and players can base their decision
on a finite number $M$ of past decisions.

Global efficiency is measured by the
standard deviation of the sum of individual decision,
\be
\sigma^2 = \left \langle \left (\sum_{i=1}^N
    \sigma_i^t\right)^2 \right \rangle_t. \label{EMG-ssqdef}
\ee

Random guessing by all players leads to $\sigma^2=N$;
a smaller value indicates good coordination among
players, a larger value is a sign of herd behavior.

One obvious feature of this game is that there is no 
unique best action for the players -- if it existed, it
would be the same for all players for symmetry reasons,
and all players would lose. 

From the point of view of economic game theory
\cite{Holler:Einfuehrung}, the game
has a large number of Nash equilibria -- combinations of
strategies where no player can improve his chances of 
winning by unilaterally changing his strategy. For example,
if $(N+1)/2$ players choose $+1$ all the time, and $(N-1)/2$
pick $-1$, the global loss is minimal ($\sigma^2=1$); 
however, those who are on the losing side stay there 
forever, and a player who switches sides will cause the 
majority to flip, and continue losing. Simple combinatorics
show that there are $\tbinom{N}{(N-1)/2}$ such combinations.

Furthermore, there are even more Nash equilibria in 
mixed strategies: e.g., if all players pick $+1$ with a 
probability of $0.5$ and $-1$ otherwise, no single player
who develops a preference for one option will have an 
advantage from this. However, if all players continue 
guessing, one gets $\sigma^2=N$, as pointed out before;
a better coordination would be desirable. A vast continuum
of mixed strategies exists where no outcome is preferred --
all of these are Nash equilibria.

In the absence of a unique best way to proceed, players
have no choice but to adapt their strategy to their
environment, i.e., the behavior of their co-players.
The MG has become a testing ground for various 
forms of ``bounded rationality'', i.e., more or less
simplistic decision and learning algorithms for
the agents, ranging from a choice between a small number
of Boolean functions \cite{Challet:Emerg.,Challet:Phase,Marsili:Exact} 
and simple neural networks 
\cite{Metzler:Interact} to evolutionary algorithms 
\cite{Johnson:Segregation}.
 
This paper presents new aspects of the Evolutionary
Minority Game (EMG) studied by Johnson et al. in \cite{Johnson:Segregation}, 
and introduces an evolutionary variation of the
stochastic strategy described in \cite{Reents:Stochastic}.
Two central questions are: 1.) what are the consequences
if a player who has ``died'' in the evolutionary process 
is replaced by a modified copy of a different player,
rather than picking a strategy at random? 2.) Can 
the prescription be generalized to more than two 
choices -- $1,\dots,Q$ instead of $\pm 1$, as suggested 
in Ref. \cite{Ein-Dor:Multi-Choice}? Let us start with a look at the original 
evolutionary MG.

\section{The original Evolutionary MG}
\label{SEC-OEMG}
In its original formulation \cite{Johnson:Segregation}, the
EMG works as follows: each player has access to a 
table which records, for each possible combination
of $M$ consecutive minority decisions, what the 
minority decision following the last occurrence of 
that combination was. Players have only two
individual features, namely a score $s_i$ and a 
probability $p_i$.  
With this probability $p_i$, they choose the 
action in the history table corresponding to the
current history; otherwise they choose the opposite
action. 

Players who win gain a point on their score, whereas
the others lose a point. If the score of a player
drops below a certain threshold $-d<0$, the player
is replaced by a new one with a reset score $s_i=0$ and a 
probability $p_i$ that is either a modified copy of his
predecessor's value or chosen entirely at random.

As was pointed out before \cite{Burgos:Selforg.}, this scheme can 
be simplified by exploiting the fact that the 
entries in the history table are decoupled, and 
that there is complete symmetry between the actions $+1$
and $-1$. The simplest interpretation that gives the
same stationary distribution $P(p)$ would therefore
be that each player picks $+1$ with probability $p_i$
and $-1$ otherwise. This points to an analogy with 
classical game theory \cite{Neumann:Theory}: 
the Minority Game is an $N$-player
negative-sum matrix game, and the $p_i$ define the 
mixed strategies of each player. 

In the original EMG, a deceased player is replaced by a new player
whose strategy $p_i$ is a modified version of his
predecessor's value: a random number  $\Delta$ with a given
variance $V$ is
added to the previous value, with reflecting or cyclic boundary
conditions at $p=0$ and $p=1$. It turns out that neither the
exact value of the threshold $d$ nor the number $N$ of players  play a
significant role, and that the typical size of mutations
changes the results quantitatively, but not qualitatively: 
the stationary probability
distribution $P(p)$ develops two peaks at $p=0$ and $p=1$,
while there is still a significant probability for
intermediate values of $p$. 

First attempts to calculate the probability distribution
analytically were only moderately accurate 
\cite{Lo:Theory,DHulst:Hamming}:
they assumed that the reason for the self-organized
segregation was only in the self-interaction of agents, 
and none of the two choices was systematically preferred.
As newer studies
\cite{Hod:Self-Segregation,Nakar:Semianalytical} have shown,
this is not true: most of the time, there is a significant
preference for one of the two options, and players who
prefer this option have higher losses and a higher chance to
be replaced. The preference for one side undergoes rather
regular oscillations, with accompanying oscillations of the
scores of players with one or the other preference. (The
presence of these oscillations also means that the
distribution $P(p)$ is time-dependent, and becomes
stationary only when averaged over a long time compared to the
oscillation period. Whenever we speak of a stationary
distribution from now on, we mean it in that sense.)

These publications also reported that, if a winning player
receives $R<1$ points rather than 1 point, the segregation
into extreme opinions vanishes below a certain point $R_c$  
and is replaced by a 
preference for undecided agents with $p\approx 0.5$.
This is a rather remarkable result: after all, the aim of
the game is unaffected by the modified payoff $R$ -- it is
still advantageous to be in the minority, the chances of
winning still only depend on the set of $\{p_i\}$, and the optimal 
configuration still has $(N-1)/2$ players on one side and
$(N+1)/2$ on the other. What has changed is the dynamics of
the game, and it is these dynamics that prevent the system
from finding a more advantageous state. 

The crucial point about the evolutionary dynamics that have
been considered so far is this: {\em they do not
  systematically favor mutations with a higher fitness} --
they are not {\em compatible} in the sense of Ref. 
\cite{Friedman:Evolutionary}: 
``\dots for any dynamic compatible with a properly defined
fitness function fitter strategies should increase relative
to less fit strategies.'' 
We will demonstrate this in the limit of infinitely large
mutations, which amounts to the same as choosing a new
strategy completely at random. Simulations indicate that 
the results apply to small mutations as well.

Let us take the fitness $f(p)$ associated with a strategy
$p$ to be the negative of the probability of a player using
$p$ of dying in a given round (in previous studies,
this was assumed to be the average gain divided by the threshold). 
The strategies are distributed
following a probability density $P(p)$, with a cumulative
probability distribution 
\be
C(p,t) = \int_0^p P(p') dp'.
\ee
The expected number of players  with $p_i<p$ who mutate in a given round
is $\int_0^p -f(p')P(p') dp'$. Out of the replacements for
these players, a fraction of $p$ has a strategy $p_i<p$. 
The updated probability function is therefore
\bea
C(p, t+1) &=& C(p, t) + \int_0^p f(p') P(p') dp' \nonumber 
\\
&&~~~~- p \int_0^1 f(p') P(p') dp'.   \label{EMG-Cupdate}
\eea
Going to continuous time and differentiating with respect to
$p$, the integro-differential equation for $P(p)$ looks as
follows:
\be
\frac{dP(p)}{dt} = P(p)f(p) - \int_0^1 P(p')f(p') dp' =
P(p)f(p) - \bar{f}. \label{EMG-dPdt1}
\ee
Keeping in mind that the fitness always takes negative values,
Eq. (\ref{EMG-dPdt1}) means that if $P(p)$ is small enough, 
it increases even if the fitness associated with is is below
average -- this is clearly not what is desired.

The problem can be remedied (in principle) by a small change
in the dynamics: a player is replaced not by a copy of
himself, but by a copy of another player. This makes sense
in various interpretations: in an economic situation, 
``dying'' could have the meaning
of ``going broke'', and a player who tries a new start wants
to imitate one of his (apparently more successful)
competitors. In a biological setting, an organism  
literally dies, and an offspring of another
organism takes its place. 
With this new mechanism, Eq. (\ref{EMG-Cupdate})
takes the form
\bea
C(p,t+1) &=& C(p, t) + \int_0^p f(p')P(p')dp' - \nonumber \\ 
&&~~~~C(p) \int_0^1 f(p') P(p')dp',
\eea
which leads to the dynamic 
\be
\frac{\dot{P}(p)}{P(p)} = f(p) - \bar{f}.
\label{EMG-dPdt2}
\ee
This describes a so-called Malthus process, where the
frequency of strategies with above-average fitness increases
exponentially. The problem with this dynamic is that once a
strategy becomes extinct, there is no way of reviving it.
In the limit of infinitely many players that was tacitly
assumed above, this is not a problem, since the probabilities for a
strategy never go to 0 in finite time.
However, with a finite number of players imitating each other, this
would eventually lead to a small number of sub-populations,
each of which exclusively plays one of the mixed strategies that
happened to survive the initial stage (this scenario
resembles a variation of the Backgammon model \cite{Godreche:Entropy}).

\begin{figure}[h]  
\epsfxsize= 0.95\columnwidth
  \epsffile{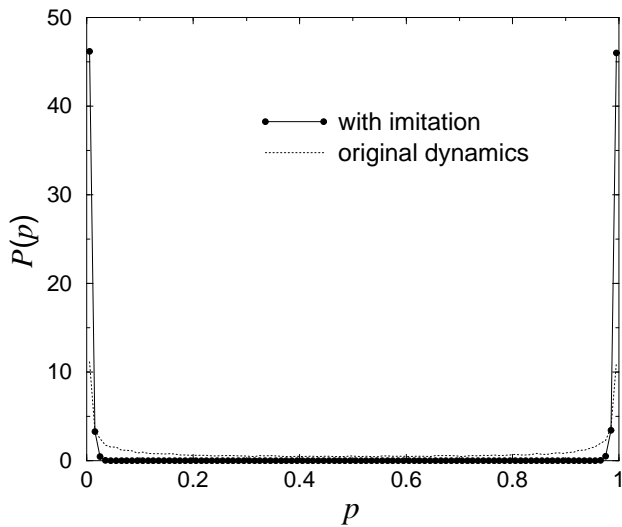}
  \caption{Stationary distribution $P(p)$ of the EMG with
    imitation, compared to the EMG without imitation with
    large mutations. Parameters were $N=501$, $d=10$, $R=1$,
    $V=10^{-4}$.} 
  \label{EMG-imit1}
\end{figure}

To get a well-defined final state independent of initial conditions, 
it is therefore necessary to add a small mutation
to the copied strategy to explore unoccupied areas in
strategy space -- for example by adding a Gaussian of
variance $V\ll 1$ to $p$, with reflecting boundary conditions.
If this is done, a stationary probability distribution
emerges that is strongly peaked at $0$ and $1$ and vanishes
for intermediate $p$, as seen in
Fig. \ref{EMG-imit1}. Global losses are reduced
dramatically:
for example, for $V= 10^{-4}$, one gets $\sigma^2 \approx
0.021N$ instead of $\sigma^2 \approx 0.31 N$ for the
original EMG for large $d$. Coordination can be improved even more by
decreasing $V$, at the cost of a longer equilibration time.
Furthermore, the results of decreasing the reward $R$ for
winning, as suggested in Ref. \cite{Hod:Self-Segregation}, 
are less dramatic: decreasing $R$ does not destroy
segregation; however, the peaks at extreme strategies become
wider, and global efficiency decreases somewhat.

\section{Multi-choice EMG}
\label{SEC-MCEMG}
Generalizations of the MG to more than two options were
studied in \cite{Ein-Dor:Multi-Choice} (with agents using neural networks) and
\cite{Chow:Multiple,Chow:Multichoice}  (with agents using a
set of decision tables). The basic idea is simple: 
\begin{itemize}
\item each player now picks an action (or ``room'')
$\sigma_i^t \in {1,\dots,Q}$ out of $Q$ options;
\item the number of players $N_q$ who chose each option
is determined: $N_q = \sum_i^N \delta_{\sigma_i,q}$;
\item the option chosen by the fewest players 
(the ``least crowded room'', with occupation $N_{min}$)
is declared winner (a coin toss decides in case of a tie);
\item the
players who chose the winning room receive an award (let us
say, one point), whereas the others lose $1/(Q-1)$ points. 
\end{itemize}
Global efficiency can be measured by taking the
analog of Eq. (\ref{EMG-ssqdef}) either for the occupation
of the winning room: 
\be
\sigma^2_{min} = \left \langle \left ( N_{min} - \frac{N}{Q}
    \right)^2 \right \rangle_t \label{EMG-ssqmdev}
\ee
or the occupation of all rooms:
\be
\sigma^2_{Q} = \frac{1}{Q} \left \langle \sum_q^Q \left ( N_{q} - \frac{N}{Q}
    \right)^2 \right \rangle_t. \label{EMG-ssqQdev}
\ee
In many cases, the latter quantity differs from $\sigma^2_{min}$ only by a
constant factor and is easier to calculate
\cite{Ein-Dor:Multi-Choice}. For the reference case of
random guessing, $\sigma^2_Q$ takes the value of $N/Q$.

The generalization of the evolutionary MG to multiple
choices leaves several options. We choose the one that
yields a standard multi-player matrix game:
each player is equipped with a strategy vector $\v{p}_i$,
with entries $p_{i,q} \geq 0$ that give the probability of
player $i$ choosing room $q$. These vectors obey the normalization
constraint $\sum_q p_{i,q} = 1$. 

How strongly a player specializes in one option can be
measured using the self-overlap of his strategy vector:
\be
\R_i = \sum_q p_{i,q}^2.
\ee
This quantity varies from $1/Q$ for a completely undecided
player to 1 for a player who chooses one option exclusively.
The average over the population, $\R = \sum_i \R_i/N$, is
therefore a good measure of the degree of specialization 
among players.

If a player's score drops below the threshold $-d$, he 
is replaced by a player with a different strategy vector. 
Again, two different paths suggest themselves: 
either, as in Ref. \cite{Johnson:Segregation}, the player is replaced by an
altered copy of himself (or, in the extreme case of 
large mutations, a randomly chosen new player); or the gap
is filled by the mutated offspring of another, randomly
chosen player. 
We ran simulations starting with random initial vectors
uniformly distributed on the simplex (see the Appendix).
The same picture emerges as in the binary-choice case:

If deceased players are replaced by copies of themselves or 
players with random strategies, very little specialization
occurs. The stationary distribution of $\R_i$ gets slightly
more contributions from larger values; however, the mean
value shifts only little. E.g., for $Q=3$ and $d=10$, 
the average $\R$ changes from $1/2$ for no adaption 
to $\approx 0.540$ for replacing deceased players 
with randomly chosen strategies.
Correspondingly, global efficiency increases only slightly:
in this case, from $\sigma^2_Q= 1/3$ to $\approx0.242$.

\begin{figure}[h]  
\epsfxsize= 0.95\columnwidth
  \epsffile{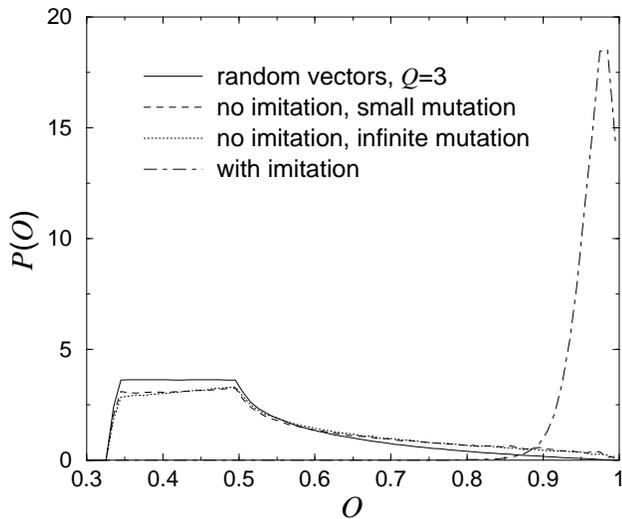}
  \caption{Stationary distribution $P(\R)$ of the
    self-overlap of strategy vectors for $Q=3$ options. 
    The solid line give the initial state of vectors chosen
    uniformly on the simplex. The dashed and dotted line
    give the result of the original dynamics: it matters
    little if replaced players undergo small mutations or
    are chosen completely at random. The dot-dashed line
    shows the effect of imitation with small mutation. 
    The same parameters as in
    Fig. \ref{EMG-imit1} were used.} 
  \label{EMG-imitQ3}
\end{figure}

However, copying another player (with small mutations) gives
excellent coordination: in the stationary state, all players
specialize strongly in one of the choices -- the
self-overlap of strategies is close to 1 (see
Fig. \ref{EMG-imitQ3}). 
The width of the
probability distribution of $\R_i$ again depends on the magnitude
of mutations: the smaller the mutations, the narrower the
peak. As before, eliminating mutations altogether prevents
coordination.

\section{The Stochastic MG} 
\label{SEC-SMG}
In Ref. \cite{Reents:Stochastic}, a strategy was presented
that looks similar to the EMG described above: again, each
agent is equipped with a probability $p_i$ that
characterizes his behavior. The meaning of $p$ is different,
though: if a player wins in a given round, he is content and
repeats his choice $\sigma_i^t$ in the following round. If he loses,
however, the agent may rethink his game plan and switch to the
opposite action with probability $p_i$. 
This prescription amounts to a one-step
Markov process which can be solved analytically in some
regimes if all players use the
same $p$ \cite{Reents:Stochastic,Horn:Diplom}.  

For large
$p$ (of order $1$), a finite fraction of the population
switches at every time step, resulting in large global
losses ($\sigma^2 = \mathcal{O}(N^2)$). Furthermore, the
majority flips very frequently. The stationary probability
distribution of $A=\sum_i {\sigma^2_i}$ 
takes the shape of two roughly Gaussian peaks centered at 
$\pm N p/(2-p)$.e
Pine finished -- Clo

However, if $p$ 
scales with $p=2x/N$, $x=\mathcal{O}(1)$, 
there is very good coordination ($\sigma^2 = 1 + 4x +
4x^2/3$ as $N \rightarrow \infty$), and the minority does
not switch at every time step. The stationary probability
distribution is centered at $A=0$, with a width of roughly
$2x$. 

The relative simplicity of the mathematical description
breaks down if each player is allowed to have an individual
$p_i$: if players are distinguishable, it is no longer
enough to state how many of them are on one side or the
other to completely characterize the system. However, with a
few approximations, even then some insight can be gained
if complications like an evolutionary dynamics are
introduced.

We start with a scenario analogous to that described in
Sec. \ref{SEC-OEMG}, which we will call SEMG (stochastic
evolutionary Minority Game): each of the $N$ players has an
individual probability $p_i$ of switching, which is
initially a uniform random number. In the case of a loss,
the player loses a point; otherwise he wins one. If his
score drops below $-d$, his probability is replaced, and his
score reset to $0$. One 
would expect that players with smaller $p$ have an advantage
over those with large $p$, and one would hope that players organize
themselves to a stationary state with $p_i\propto 1/N$.

\begin{figure}[h]  
\epsfxsize= 0.95\columnwidth
  \epsffile{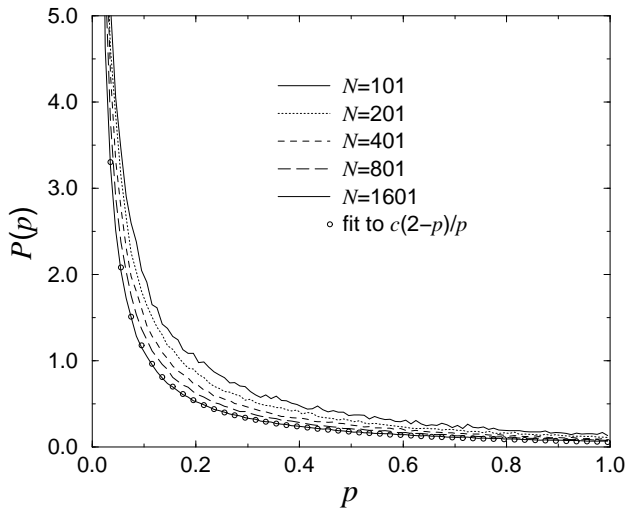}
  \caption{Stationary distribution $P(p)$ in the Stochastic
    EMG, for different values of $N$. New players use
    randomly chosen values of $p$; the threshold is $d=10$.} 
  \label{EMG-st1}
\end{figure}

If mutations of players are performed analogous to the
original EMG (new players are modified copies of the
deceased ones), a stationary distribution $P(p)$
emerges in which small $p$ are more likely than large ones,
but there is still a significant tail towards large $p$
(see Fig. \ref{EMG-st1}). 
Simulations show that contrary to the 
original EMG, the details of the
mutation process (size of the mutation, reflecting/ cyclical
boundary conditions etc.) have no impact at all on this
distribution. Neither does the threshold $d$.
The number of players $N$ has only a small effect -- 
the shape of the distribution for larger $p$ does not
change significantly, but $P(p)$ increases for very small $p$ as $N$
increases. This effect is explained later; however, it
is much too small to achieve a mean $p$ that scales with
$1/N$. 

As mentioned before, $p=\mathcal{O}(1)$ means that the
minority changes sides at practically every time
step. Assuming that this is true, it is possible to
calculate the average gain of a player with a given $p$,
and hence his expected lifetime. From this, the stationary
distribution can be calculated. Let us start with a more
general approach, which assumes very simple dynamics of the
global decision and neglects the impact of a single player
on that decision. If the global minority stays the same in
two consecutive time steps with a given
probability $\mu$, the average gain of a player can be
determined as follows: with a probability $w(p)$, the player
wins in a given round. Consequently, he does not switch
sides, and wins again in the next round with probability
$\mu$. However, if he lost in the first round, he can win by
changing sides if the minority stays on the same side
(probability $\mu p$) or if he insists on his opinion, and
the minority switches (probability $(1-\mu)(1-p)$). 
For a given $\mu$, the probability of winning is therefore
\be
w(p) = \mu w(p) + (\mu p + (1-\mu)(1-p))(1-w(p)). 
\ee
For the average gain $g(p)= 2w(p)-1$, this gives
\be
g(p) = \frac{ - p(1-2\mu)}{1+(1-p)(1-2\mu)}.
\label{EMG-g_p}
\ee
As long as $g(p)$ is systematically negative and does not
undergo fluctuations on time scales comparable to the
lifetime of a player, it is safe to assume that the mean
lifetime $L(p)$ is  $d/g(p)$ -- the score is a random walk with a
negative drift which outweighs the diffusive motion for
sufficiently large $d$. 

Assuming that the average $p$ is large enough to ensure that
the majority will flip at every time step -- i.e., $\mu=0$,
and $g(p)= p /(p-2)$ -- one can now identify $-1/L(p)=p/d(2-p)$ with
the fitness $f(p)$ and use Eq. (\ref{EMG-dPdt1}) to derive,
for the stationary state,
\be
P(p) \propto \frac{2-p}{p}. \label{EMG-semgstat1}
\ee   
This probability distribution has the unpleasant property of
diverging at $p=0$, since agents with $p=0$ are assigned an
infinite lifetime. In practice, three effects come into
play: first, agents never have exactly $p=0$. 
Second, even for agents with $p=0$, their impact on the decision
will give them a very small negative average gain and a
long, but not infinitely long life. Third,  
$\mu$ is very small, but never strictly equal to $0$; 
if the probability of large values of $p$ becomes too small,
the simplistic assumptions about the dynamics no longer
hold, and $\mu$ increases.Together, these effects 
are responsible that for any given set of parameters, a
stationary distribution emerges.
Eq. (\ref{EMG-semgstat1}) allows for a good fit to these stationary
distributions measured in simulations, as seen in
Fig. \ref{EMG-st1}. 

The average $p$ that emerges from these simulations is of
order $1$ (to be more specific, around $0.10$, with the
precise value depending on $N$ and $d$). This means that the
solution is self-consistent: the value of $p$ that results
from the dynamics is large enough to justify the assumptions
that went into estimating it.

Analogous to Sec. \ref{SEC-OEMG}, the evolutionary dynamic
of replacing a player with a random player or a copy of the
old one does not always favor strategies with higher
fitness. However, the same step can be taken to improve
coordination: if deceased players are replaced with a copy
of another player chosen at random, the relative growth
of $P(p)$ is proportional to $f(p)- \bar{f}$, just as in
Eq. (\ref{EMG-dPdt2}).

Unfortunately, Eq. (\ref{EMG-g_p}) is not applicable for very
small $p$ (since it neglects the impact of the considered
agent), and there is no simple equation that gives
the fitness as a function of the strategy for all
regimes. Nevertheless, there seems to be no situation where
having a higher $p$ gives better results. Hence, the
evolutionary dynamic should lead to a state with minimal $p$
for all agents. A similar problem as in Sec.  \ref{SEC-OEMG}
comes into play here, although it does not have quite as
troubling effects: with a finite number of players, the
best possible coordination is for all players to adapt the
smallest value of $p$ that survived the initial
stage. However, this value is usually of order $1/N$ -- if
initial values are chosen at random, they have an average
distance of $1/N$.

\begin{figure}[h]  
\epsfxsize= 0.95\columnwidth
  \epsffile{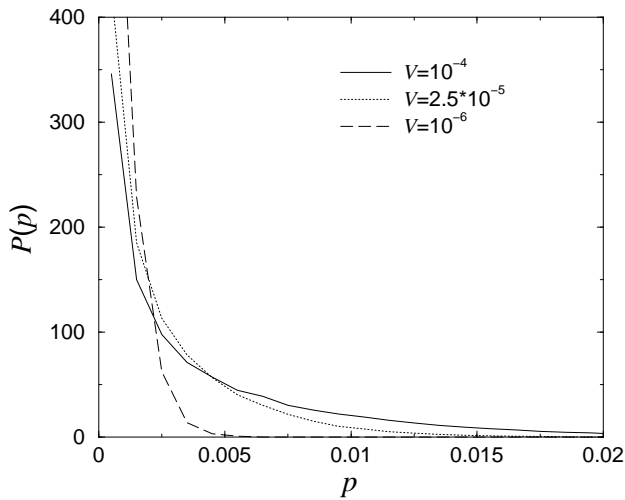}
  \caption{Stationary distribution $P(p)$ in the Stochastic
    EMG with copying. The variance $V$ of mutations 
    influences the width of the distribution. 
    The simulation included $N=2000$ players, with a 
    threshold $d=10$.} 
  \label{EMG-st2}
\end{figure}

Just as in Sec. \ref{SEC-OEMG}, the sensitivity to 
the initial state can be removed by adding a small random
number to $p_i$ when a new player is created. As seen in
Fig. \ref{EMG-st2}, results are
similar: a peak centered at $p=0$ emerges, whose width
depends on the size of the mutations. With sufficiently
small mutations, $p=\mathcal{O}(1/N)$ and $\sigma^2=
\mathcal{O}(1)$ can easily be achieved. 

One of the drawbacks of introducing small mutations is that
their size is a new parameter that has to be adjusted to get
a $\sigma^2$ of order 1. One of the conceptual 
flaws of the SMG was that players had to be aware of the 
size of the population to justify an adequate choice of $p$. 
One might have hoped that in an evolutionary scheme, the
correct scaling for $p$ would emerge naturally. With the
dynamic of copying with mutations, it does not. Maybe an
evolutionary mechanism that mutates the size of mutations
would solve this.

\section{Multi-choice stochastic MG} 
The stochastic MG can be generalized to multiple choices in
several ways, none of which is quite as natural as the
generalization of the EMG in Sec. \ref{SEC-MCEMG}. One of
the most intuitive ways is this: again, players have the
choice between $Q$ different actions, or ``rooms''. Those who chose the
least-crowded room are content and choose it again,
whereas all others decide, with probability $p_i$, to look
for alternatives. If they are not informed about which
room won, they  randomly pick one of the room that 
they did not choose the last time. Another plausible scenario is that
they know which room won, and choose it with
probability $p$ the next time.

In both cases, the system can still be considered a one-step 
Markov process. However, the state of the system must now be
characterized by $Q-1$ values $N_1, \dots, N_{Q-1}$,
which give the number of players that chose each action (the
remaining value $N_Q$ can be calculated from normalization
constraints: $\sum_q N_q =N$), and the joint probability
distribution is a $Q-1$-dimensional tensor, or a function
living in $\mathbb{R}^{Q-1}$ if one wants to go to continuous  
variables. Transition probabilities look even worse, taking the
form of $2(Q-1)$-dimensional tensors or integral kernels.
Put briefly, this problem is only accessible to simulations
and very crude approximations. 

\begin{figure}[h]
\epsfxsize= 0.95 \columnwidth
\centerline{
   \epsffile{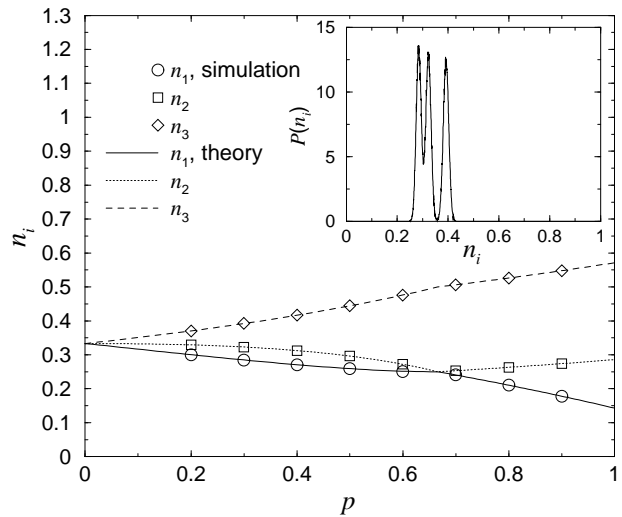}}
  \caption{Centers of the peaks of $\pi(n)$ in the
    stochastic MG with $Q=3$. Simulations 
    agree well with Eqs. (\ref{EMG-q3sol3}). The inset shows
    the probability distribution for $p=0.3$ and $N=2000$.}
  \label{EMG-pottsq3ns}
\end{figure}

In the limit of $p = \mathcal{O}(1)$, $N\rightarrow \infty$,
the behavior is analogous to that detailed in
Sec. \ref{SEC-SMG}: finite fractions of players
move from room to room, and the minority option changes 
at every time step. Suitable variables are $n_q= N_q/N$, 
the fractions of players who chose option $q$.
Occupation probabilities $P(n_q)$  
turn out to be a superposition of $Q$ Gaussian peaks whose widths
decrease with $N$. Self-consistent values for the centers 
of the peaks can be found analytically, as the following
example for $Q=3$ will show:

 At any given step, there are three occupation
numbers, which we order $n_1 < n_2< n_3$.
Room 1 will now receive players from rooms 2 and 3, whereas
rooms 2 and 3 gain players from the respective other room 
and lose players to all other rooms. Neglecting
fluctuations, the rate equations for the 
occupations $n_i^+$ at the next time step look like this:
\bea
n_1^+ &=& n_1 + (p/2) (n_2 + n_3); \nonumber \\
n_2^+ &=& n_2  -p n_2 + (p/2) n_3; \nonumber \\
n_3^+ &=& n_3  -p n_3 + (p/2) n_2. \label{STO-q3rates}
\eea
If one can find a permutation of $n_i$ such that each 
$n_i$ is equal to $n_j^+$ with some $j\neq i$, one 
has a solution. In the present case, the solution 
for Eq. (\ref{STO-q3rates}) is 
$n_1^+ = n_3$, $n_2^+ = n_1$ 
and $n_3^+ = n_2$ for $0<p\leq 2/3$,
and  $n_1^+ = n_3$, $n_2^+ = n_2$ and $n_3^+ = n_1$
for $2/3 \leq p <1$. The corresponding equations are
\bea 
n_1 &=& \frac{4-6p+3p^2}{3(2-p)^2} \mbox{ for } p\leq2/3, 
   1- \frac{6}{10-3p} \mbox { for } p>2/3; \nonumber \\
n_2 &=& \frac{4(1-p)}{3(2-p)^2} \mbox{ for } p\leq2/3, 
   \frac{2}{10-3p}    \mbox { for } p>2/3;  \nonumber \\
n_3 &=& \frac{2}{3(2-p)} \mbox{ for } p\leq2/3, 
    \frac{4}{10-3p}  
    \mbox { for } p>2/3.
\label{EMG-q3sol3}
\eea
This solution agrees well with behavior observed in
simulations, as Fig. \ref{EMG-pottsq3ns} shows.

\begin{figure}[h]
\epsfxsize= 0.95 \columnwidth
\centerline{
   \epsffile{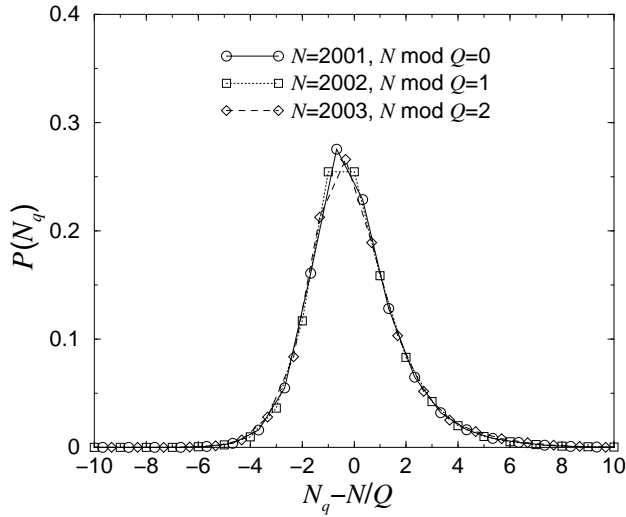}}
  \caption{Stationary probability distribution of occupation
    numbers $N_q$ in the multi-choice SMG, with uniform
    $p=1/2000\approx 1/N$, for $Q=3$.}
  \label{SMG-Q3dist}
\end{figure}

For small uniform $p$ of order $Q/N$, analytical results are
hard to find, for the mentioned reasons. Evidence from
simulations shows that the system organizes itself into a
probability distribution close to optimal coordination. The
details of the distribution depend, even for large $N$, on 
$N \bmod Q$. An example of such a distribution is shown in
Fig. \ref{SMG-Q3dist}.

Evolutionary dynamics can be introduced exactly analogous to
the previous sections, and very similar results are observed:
choosing a new player at random and replacing a player by a
mutated copy of himself yields the same stationary
probability distribution, which has a long tail  towards
larger $p$. Average values for $p$ are around $0.2$, the
probability distribution of occupation numbers $N_q$ shows
multiple peaks.

The alternative dynamic of copying another player with
mutation gives a sharp peak around $p=0$, with $\sigma^2_Q$
on the order of 1 for sufficiently small mutations.

\section{Concluding remarks} 
We have shown that the self-organized segregation observed
in the evolutionary Minority Game is not only much more
pronounced, but also more robust to modifications of the
payoff scheme if a suitable dynamic is used -- one that
allows strategies with above-average fitness to grow, rather
than keeping sub-par strategies alive. Copying another
player's strategy is a suitable way of doing this. The copy
has to be modified by a small mutation to eliminate sensitivity 
to initial conditions. 

We have also introduced a natural generalization of the EMG
to multiple choices, evolutionary dynamics for the
Stochastic MG, and an extension of the latter to multiple
choices. The results in all cases have striking
similarities: properly chosen evolutionary dynamics lead to 
near-optimal coordination and drastic suppression of losses,
compared to random guessing. 

It is somewhat ironic that the key ingredient to these
dynamics is copying the strategy of another player -- in a game
where the goal is to take different actions than the majority.
One could argue that in the regime of the EMG where players have very
strong preferences for either side, the EMG is not too
different from the Stochastic MG: players stick to their
opinion if possible; if they lose too often, they either
copy a player from their own side (which changes little
about the situation), or they copy a player from the other
side -- they switch their output. The presence of
accumulated scores makes the situation more complicated than
that, yielding preference oscillations whose wavelength
depends on the threshold. 

Despite the increased efficiency and robustness that the  
imitation mechanism has brought, the dynamics are still
complex, and a thorough analytical treatment has not been
found yet. The same holds for the SEMS, where the 
interplay between the probability distributions 
of outputs and of strategies is difficult to handle 
analytically. Maybe future research will fill in the missing
details.

\begin{acknowledgments}
R. M. acknowledges financial support by the GIF.
We thank Georg Reents for his enthusiasm about the
Stochastic MG, for ideas, helpful discussions and
supervision. We also thank Wolfgang Kinzel for his support
and supervision.
\end{acknowledgments}

\appendix 
\section{Constructing uniformly distributed  vectors on a simplex}
There are many conceivable methods of finding
$Q$-dimensional random vectors $\v{p}$ that obey 
the constraints of probabilities, $p_i\geq 0$ and 
$\sum_i p_i = 1$. However, the easiest ones do not give a
uniform distribution on the simplex of allowed vectors:
for example, forcing a set of uniform random numbers between
0 and 1 to obey the constraints by dividing them by 
their sum emphasizes vectors in
the center of the simplex due to projection effects.

The following method generates uniformly distributed 
probability vectors from uniformly
generated random numbers. We include it because it may be
useful to the reader for other applications.

The space of allowed vectors $\v{p}$ is spanned by 
linear combinations
\be
\v{p} = \sum_q a_q \v{b}_q
\label{EMG-lincomb}
\ee
of a set of $Q$ basis  vectors $\v{b}_q$
\be
\v{b}_1 = \left(\begin{array}{r} 1 \\0 \\0 \\ \vdots
  \end{array} \right),
\v{b}_2 = \left(\begin{array}{r} -1 \\1 \\0 \\ \vdots
  \end{array} \right),
\v{b}_3 = \left(\begin{array}{r} 0 \\-1 \\1 \\ \vdots
  \end{array} \right), \dots
\ee
with coefficients $1=a_1\geq a_2\geq\dots \geq a_Q \geq 0$. 
The components must be chosen
with a suitably weighted probability distribution to account
for the fact that a larger coefficient $a_2$ allows for more
combinations of $a_3$, $a_4$ etc. Since the volume of the
sub-simplex limited by $a_q$ is proportional to
$a_q^{Q-q}$, the appropriate distribution is
\be
\mbox{Prob}(a_q) = 
\left \{ \begin{array}{ll} \frac{Q-q+1}{a_{q-1}} a_q^{Q-q}
      &\mbox{~~~for   } 0 \leq a_q \leq a_{q-1} \\  0 &
      \mbox{~~~else} 
\end{array} \right. .
\ee
Consequently, a set of coefficients $\{a_2, \dots, a_Q \}$ can
be calculated from a set of random numbers $\{r_2,
\dots,r_Q\}$ uniformly distributed between $0$ and $1$ by 
a simple transformation \cite{Press:NumRec}:
\be
a_1 = 1; \ \ a_q = a_{q-1}  r_q^{1/(Q-q+1)}.
\ee
Eq. (\ref{EMG-lincomb}) then gives the desired vector on the simplex.

\ \\

\end{document}